\documentclass[epj]{svjour}
\usepackage{amscd}
\usepackage{amssymb}
\usepackage{amsmath}
\usepackage{dcolumn}
\usepackage[T1]{fontenc}
\usepackage[latin1]{inputenc}
\usepackage{graphicx}
\usepackage{graphics}
\usepackage{color}
\usepackage{latexsym}
\usepackage{amsfonts}
\def\be{\begin{equation}}
\def\ee{\end{equation}}
\def\ba{\begin{eqnarray}}
\def\ea{\end{eqnarray}}
\def\>{\rangle}
\def\<{\langle}
\def\n{\nonumber}

\def\sc{\scriptsize}

\begin{document}
\title{The second law of thermodynamics
       in the quantum Brownian oscillator at an arbitrary temperature}
\author{ILki Kim\inst{1}\thanks{\emph{e-mail}: hannibal.ikim@gmail.com}
        \and G\"{u}nter Mahler\inst{2}}
\institute{Division of Natural Science and Mathematics, St.
           Augustine's College,
           Raleigh, NC 27610, U.S.A. \and
           Institute of Theoretical Physics I,
           University of Stuttgart, 70550 Stuttgart, Germany}
\date{\today}
%
\abstract{In the classical limit no work is needed to couple a
system to a bath with sufficiently weak coupling strength ({\em or}
with arbitrarily finite coupling strength for a linear system) at
the same temperature. In the quantum domain this may be expected to
change due to system-bath entanglement. Here we show analytically
that the work needed to couple a single linear oscillator with
finite strength to a bath cannot be less than the work obtainable
from the oscillator when it decouples from the bath. Therefore, the
quantum second law holds for an arbitrary temperature. This is a
generalization of the previous results for zero temperature
\cite{FOR06,KIM06}; in the high temperature limit we recover the
classical behavior.
\PACS{
      {03.65.Ud}{Entanglement and quantum nonlocality}   \and
      {05.40.-a}{Fluctuation phenomena, random processes, noise, and Brownian motion}   \and
      {05.70.-a}{Thermodynamics}
     } 
}
\authorrunning{ILki Kim and G\"{u}nter Mahler}
\titlerunning{The second law of thermodynamics in the quantum Brownian oscillator $\cdots$}
\maketitle
%
\section{Introduction}
The second law of thermodynamics \cite{CAL85} is considered one of
the central laws of science, engineering, and technology. For over a
century it has been assumed to be inviolable by the scientific
community. Over the last 20 years, however, its absolute status has
come under increased scrutiny \cite{CAP05}. Challenges to the second
law have recently attracted big interest with consideration of the
miniaturization of a system under investigation, especially at low
temperatures where quantum effects are important \cite{CAP05,SPI05}.
In contrast to common quantum statistical mechanics which is
intrinsically based on a vanishingly small coupling between system
and bath (``thermodynamic limit''), the finite coupling strength
between them in the quantum regime causes some subtleties that must
be recognized. The {\em quantum} thermodynamic behaviors of small
systems have theoretically been investigated intensively and
extensively \cite{FOR06,KIM06,CAP05,SPI05,MAH04,FOR05,BUT05,HAE06}
and experimentally been examined \cite{CAP05,SHE05,BER05}.

The problem of a quantum linear oscillator coupled to an
independent-oscillator model of a heat bath (quantum Brownian
motion) has been extensively discussed
\cite{FOR85,FOR88,LEW88,HAE05,GHO05}. The validity of the quantum
second law has recently been questioned in this scheme at zero
temperature \cite{SPI05,ALL00,NIE02} by the fact that the coupled
oscillator has a higher average energy value than the free harmonic
oscillator ground state, which would not be in compliance with the
second law. By means of a cyclic coupling/decoupling process one
might expect to extract useful work from a single bath. However,
this claim turned out incorrect; the apparent excess energy in the
coupled oscillator cannot be used to extract useful work, neither
for the well-known Drude damping model with a cut-off frequency for
the spectral density of bath modes shown by Ford and O'Connell in
\cite{FOR06} nor for both discrete bath modes and continuous bath
modes with the generalized realistic damping models \cite{KIM06},
since the minimum value of the work to couple the free oscillator to
a bath takes above and beyond this excess energy. Therefore, the
quantum second law for zero temperature is inviolate.

In this paper, we would like to discuss the second law in the
quantum Brownian motion at an {\em arbitrary} temperature. We will
obtain an analytic expression for the second-law inequality which
can explicitly be shown to hold for the Drude model, which is the
prototype for physically realistic damping. It is known \cite{WEI99}
that a finite frequency cut-off reflects the physical fact that the
bath cannot react instantaneously to a change of the system
oscillator, and that in the absence of the cut-off, some observables
such as the variance of the system momentum diverge. Therefore, the
physically unrealistic cutoff-free damping models considered in
\cite{KIM06} will not extensively be considered here ({\em cf}.
Sect. \ref{sec:ohmic}). From the result of this work, the appearance
({\em or} disappearance) of quantum effects versus thermal
fluctuation over the different temperatures will also be seen
explicitly. Let us begin with a brief review on the basics of the
quantum Brownian motion. We will below adopt the notations used in
\cite{ING98}.

\section{Basics of quantum Brownian motion}
The quantum Brownian motion under consideration is described by the
model Hamiltonian
\begin{equation}\label{eq:total_hamiltonian1}
    \hat{H}\; =\; \hat{H}_s\; +\; \hat{H}_{b-sb}\,,
\end{equation}
where
\begin{eqnarray}\label{eq:total_hamiltonian2}
    \hat{H}_s &=& \frac{\hat{p}^2}{2 M}\, +\,
    \frac{M}{2}\,\omega_0^2\,\hat{q}^2\\
    \hat{H}_{s-sb} &=& \sum_{j=1}^N \left\{\frac{\hat{p}_j^2}{2 m_j} +
    \frac{m_j}{2} \omega_j^2 \left(\hat{x}_j - \frac{c_j}{m_j\,\omega_j^2}\,\hat{q}\right)^2\right\}\,.
\end{eqnarray}
The Hamiltonian $\hat{H}_{s-sb}$ splits into the bath and the
coupling term,
\begin{eqnarray}
    \hat{H}_b &=& \sum_{j=1}^N \left(\frac{\hat{p}_j^2}{2 m_j} +
    \frac{m_j}{2} \omega_j^2\,\hat{x}_j^2\right)\\
    \hat{H}_{sb} &=& -\hat{q} \sum_{j=1}^N c_j\,\hat{x}_j\, +\, \hat{q}^2
    \sum_{j=1}^N \frac{c_j^2}{2 m_j\,\omega_j^2}\,.
\end{eqnarray}
From the hermiticity of Hamiltonian, the coupling constants $c_j$
are real-valued. Without any loss of generality, we assume that
\begin{equation}\label{eq:frequency_relation1}
    \omega_1\, \leq\, \omega_2\, \leq\, \cdots\, \leq\, \omega_{N-1}\, \leq\, \omega_N\,.
\end{equation}
By means of the Heisenberg equation of motion, we can derive the
quantum Langevin equation
\begin{equation}\label{eq:eq_of_motion1}
    \textstyle M\,\ddot{\hat{q}} \, +\,
    M \int_0^t d s\, \gamma(t -s)\, \dot{\hat{q}}(s)\,
    +\, M\,\omega_0^2\,\hat{q}\; =\; \hat{\xi}(t)\,,
\end{equation}
where we used $\hat{p} = M \dot{\hat{q}}$, and the damping kernel
and the noise operator are respectively given by
\begin{eqnarray}\label{eq:damping_kernel1}
    \displaystyle \gamma(t) &=& \frac{1}{M} \sum_{j=1}^N
    \frac{c_j^2}{m_j\,\omega_j^2}\cos(\omega_j\,t)\,;\,
    \displaystyle \hat{\xi}(t)\, =\, - M \gamma(t)\,\hat{q}(0) +\n\\
    && \sum_{j=1}^N c_j \left\{\hat{x}_j(0) \cos(\omega_j\,t)\, +\,
    \frac{\hat{p}_j(0)}{m_j\,\omega_j} \sin(\omega_j\,t)\right\}\,.
\end{eqnarray}
Here, $\<\hat{\xi}(t)\>_{\rho_{b'}} = 0$ and
$\<\hat{\xi}(t)\,\hat{\xi}(t')\>_{\rho_{b'}} = M \gamma(t-t')/\beta$
for the initial bath state with the shifted canonical equilibrium
distribution, $\hat{\rho}_{b'} = e^{-\beta
\hat{H}_{b-sb}(0)}/Z^{(b')}_{\beta}$ \cite{WEI99} where $\beta =
1/k_B T$, and $Z^{(b')}_{\beta}$ is the normalization constant ({\em
i.e.}, the partition function). The Fourier-Laplace transform of
$\gamma(t)$ is \cite{ING98}
\begin{equation}
    \tilde{\gamma}(\omega)\; =\; \frac{i \omega}{M} \sum_j^N \frac{c_j^2}{m_j\,\omega_j^2}\,
    \frac{1}{\omega^2 - \omega_j^2}\,.\label{eq:gamma_tilde1}
\end{equation}
Introducing the spectral density of bath modes as a characteristic
of the bath,
\begin{equation}\label{eq:spectral_density1}
    J(\omega)\; =\; \pi \sum_{j=1}^N \frac{c_j^2}{2 m_j\,
    \omega_j}\,\delta(\omega - \omega_j)\,,
\end{equation}
we can also express the damping kernel as
\begin{equation}\label{eq:damping_kernel2}
    \gamma(t)\; =\; \frac{2}{M} \int_0^{\infty} \frac{d \omega}{\pi}
    \frac{J(\omega)}{\omega} \cos(\omega\,t)\,.
\end{equation}

We now consider a response function \cite{ING98}, $\chi(t) =
\frac{i}{\hbar}\,\times$\\$\left\<[\hat{q}(t),\,\hat{q}]\right\>_{\beta}$,
where the expectation value $\<\cdots\>_{\beta}$ is taken with
respect to the equilibrium state, $\hat{\rho}_{\beta} = e^{-\beta
\hat{H}}/Z_{\beta}$ with the partition function, $Z_{\beta} =
\mbox{Tr}\,e^{-\beta \hat{H}}$. The Fourier-Laplace transform of
$\chi(t)$ is then the dynamic susceptibility
\begin{equation}\label{eq:chi_tilde1}
    \tilde{\chi}(\omega)\; =\; \frac{1}{M}\,\frac{1}{\omega_0^2 -\omega^2 -
    i \omega\,\tilde{\gamma}(\omega)}\,,
\end{equation}
which plays important roles later. It is known \cite{FOR85} that the
susceptibility $\tilde{\chi}(\omega)$ can be rewritten as
\begin{equation}\label{eq:susceptibility2}
    \tilde{\chi}(\omega)\; =\; -\frac{1}{M}\, \frac{\displaystyle
    \prod_{j=1}^N\, (\omega^2 - \omega_j^2)}{\displaystyle \prod_{k=0}^N\, (\omega^2 -
    \bar{\omega}_k^2)}\,,
\end{equation}
where the normal-mode frequencies, $\bar{\omega}_k$ of the total
system $\hat{H}$ satisfy $\omega_0^2 - \bar{\omega}_k^2 -
i\,\bar{\omega}_k\,\tilde{\gamma}(\bar{\omega}_k) = 0$. Without any
loss of generality, we here assume that
\begin{equation}\label{eq:frequency_relation2}
    \bar{\omega}_0\, \leq\, \bar{\omega}_1\, \leq\, \cdots\, \leq\,
    \bar{\omega}_{N-1}\, \leq\, \bar{\omega}_N\,.
\end{equation}
It can then be found \cite{LEW88,KIM06} that
\begin{equation}\label{eq:frequency_relation3}
    \bar{\omega}_0\, \leq\, \omega_1\, \leq\, \bar{\omega}_1\, \leq\, \cdots\,
    \leq\, \omega_{N-1}\, \leq\, \bar{\omega}_{N-1}\, \leq\, \omega_N\, \leq\,
    \bar{\omega}_N\,,
\end{equation}
and $\bar{\omega}_0 \leq \omega_0 \leq \bar{\omega}_N$.

The damping function $\tilde{\gamma}(\omega)$ in
(\ref{eq:gamma_tilde1}) can be rewritten as
\begin{eqnarray}
    \hspace{-.5cm}&&{\textstyle \tilde{\gamma}(\omega)\; =\; \frac{i}{M} \int_0^{\infty} \frac{d \omega'}{\pi}
    \frac{J(\omega')}{\omega'} \left(\frac{1}{\omega' + \omega}\, -\, \frac{1}{\omega' -
    \omega}\right)}\label{eq:damping_kernel2_1}\\
    \hspace{-.5cm}&&{\textstyle \tilde{\gamma}(\omega)\big{|}\vspace*{2.5cm}_{\omega \to \atop \omega +
    i\,0^{+}}\; =\; \frac{J(\omega)}{M\, \omega}\, +}\n\\
    \hspace{-.5cm}&&{\textstyle \hspace*{1.3cm}\frac{i}{M} \int_0^{\infty} \frac{d \omega'}{\pi}
    \frac{J(\omega')}{\omega'}\; P\left(\frac{1}{\omega' + \omega} - \frac{1}{\omega' -
    \omega}\right)\,,}\label{eq:damping_kernel3}
\end{eqnarray}
which resulted from the Fourier-Laplace transform of
(\ref{eq:damping_kernel2}). Here, we used the well-known formula,
$1/(x + i\,0^+) = P(1/x) - i \pi \delta(x)$ for $x = \omega' -
\omega$. Equation (\ref{eq:damping_kernel3}) is convenient for the
case of a continuous distribution $J(\omega)$ of bath modes; for the
simple Ohmic case, $J_{0}(\omega)\, =\, M \gamma_o\,\omega$ with an
$\omega$-independent constant $\gamma_o$, we easily have
$\gamma_0(t) = 2 \gamma_o\,\delta(t)$, and
$\tilde{\gamma}_{0}(\omega) = \gamma_o$ with a vanishing principal
({\em or} imaginary) part in (\ref{eq:damping_kernel3}), while for
the Drude model where $J_d(\omega)\, =\,
M\,\gamma_o\,\omega\,\omega_d^2/(\omega^2 + \omega_d^2)$ with a
cut-off frequency $\omega_d$, we have $\gamma_d(t) = \gamma_o\,
\omega_d\, e^{-\omega_d\,t}$, and
\begin{equation}\label{eq:drude_gamma1}
    \tilde{\gamma}_d(\omega)\; =\; \frac{\gamma_o\, \omega_d^2}{\omega^2 + \omega_d^2}\; +\;
    i\,\frac{\gamma_o\, \omega_d\, \omega}{\omega^2 + \omega_d^2} \; =\;
    \frac{\gamma_o\, \omega_d}{\omega_d - i \omega}\,.
\end{equation}
Here, $J_d(\omega)$ behaves like $J_0(\omega)$ for small frequencies
(with $\omega_d \to \infty$).

The model Hamiltonian in (\ref{eq:total_hamiltonian1}) can also
describe the classical Brownian motion (see, e.g., \cite{ZWA01}). In
considering the quantum second law below, therefore its classical
counterpart will be discussed briefly in comparison.

One might think of using instead of $\hat{H}_{sb}$ in
(\ref{eq:total_hamiltonian2}) its rotating wave approximation
$\hat{H}_{sb}^{(r)} = \hbar
\sum_j\,\kappa_j\,(\hat{a}\,\hat{b}_j^{\dagger}\, +\,
\hat{a}^{\dagger}\,\hat{b}_j)$, which has for $\omega_0 = \omega_j$
for all $j$, energy-conserving terms only. Here, $\hat{a} =
\sqrt{\frac{M \omega_0}{2 \hbar}}\,\hat{q} + \frac{i}{\sqrt{2 M
\hbar \omega_0}}\,\hat{p}$ and $\hat{b}_j = \sqrt{\frac{m_j
\omega_j}{2 \hbar}}\,\hat{x}_j + \frac{i}{\sqrt{2 m_j \hbar
\omega_j}}\,\hat{p}_j$, respectively. In this case, though, we
cannot observe any excess energy of the coupled oscillator at zero
temperature since the system oscillator and all bath oscillators
remain unchanged in their ground states, respectively, with no
entanglement (see also \cite{SEN95,KIM04}); from the Heisenberg
equation, $i \hbar\,\dot{\hat{H}}_s^{(r)} =
[\hat{H}_s,\,\hat{H}^{(r)}] = [\hat{H}_s,\,\hat{H}_{sb}^{(r)}]$, we
can obtain
\begin{eqnarray}\label{eq:derivative_of_h_s1_rwappro}
    \dot{\hat{H}}_s^{(r)} &=& \frac{\hbar \omega_0}{i} \sum_j\,\kappa_j\,\left(\hat{a}^{\dagger}\,\hat{b}_j\, -\,
    \hat{a}\,\hat{b}_j^{\dagger}\right)\n\\
    \dot{\hat{H}}_j^{(r)} &=& \frac{\hbar \omega_j}{i} \kappa_j\,\left(\hat{a}\,\hat{b}_j^{\dagger}\,
    -\, \hat{a}^{\dagger}\,\hat{b}_j\right)\,,
\end{eqnarray}
which yield $\dot{\hat{H}}_s^{(r)}|\psi_i\> =
\dot{\hat{H}}_j^{(r)}|\psi_i\> = 0$ for the initial state $|\psi_i\>
= |0\> |00 \cdots\>$, respectively. For the full Hamiltonian
$\hat{H}$ in (\ref{eq:total_hamiltonian1}), on the other hand, we
have, after a fairly lengthy calculation,
\begin{eqnarray}\label{eq:derivative_of_h_s2}
    \dot{\hat{H}}_s &=& \frac{\hbar}{2 i}
    \sqrt{\frac{\omega_0}{M}}\,
    \sum_j\,\frac{1}{\sqrt{m_j \omega_j}}\,(\hat{a}\,-\,\hat{a}^{\dagger})\,(\hat{b}_j\,+\,\hat{b}_j^{\dagger})\, +\n\\
    && \frac{\hbar}{2 i M}
    \left\{(\hat{a}^{\dagger})^2\,-\,\hat{a}^2\right\} \sum_j\,\frac{c_j^2}{m_j \omega_j^2}\,,
\end{eqnarray}
which clearly gives rise to $\dot{\hat{H}}_s |\psi_i\> \ne 0$.

\section{Formulation of the quantum second law}\label{sec:2nd_law}
From the fluctuation-dissipation theorem \cite{WEI99}, we can easily
have
\begin{eqnarray}\label{eq:dissipation_fluctuation_theorem}
    \hspace*{-.5cm}&& \textstyle \frac{1}{2}\, \<\hat{q}(t_1)\,\hat{q}(t_2)\, +\,
    \hat{q}(t_2)\,\hat{q}(t_1)\>_{\beta}\; =\\
    \hspace*{-.5cm}&& \textstyle \frac{\hbar}{\pi}\,\int_0^{\infty} d\omega\,
    \coth\left(\frac{\beta \hbar \omega}{2}\right)\,\cos\{\omega(t_2 -
    t_1)\}\; \mbox{Im}\{\tilde{\chi}(\omega + i\,0^+)\}\,,\n
\end{eqnarray}
which immediately yields
\begin{eqnarray}
    \hspace*{-.5cm}\textstyle \<\hat{q}^2\>_{\beta} &=& \textstyle \frac{\hbar}{\pi}\,\int_0^{\infty} d\omega\,
    \coth\left(\frac{\beta \hbar \omega}{2}\right)\,
    \mbox{Im}\{\tilde{\chi}(\omega +
    i\,0^+)\}\label{eq:x_correlation1}\\
    \hspace*{-.5cm}\textstyle \<\dot{\hat{q}}^2\>_{\beta} &=& \textstyle \frac{\hbar}{\pi}\,\int_0^{\infty} d\omega\,
    \omega^2\,\coth\left(\frac{\beta \hbar \omega}{2}\right)\,
    \mbox{Im}\{\tilde{\chi}(\omega + i\,0^+)\}\label{eq:x_dot_correlation1}
\end{eqnarray}
and thus the energy of the coupled oscillator
\begin{eqnarray}\label{eq:energy1}
    {\textstyle E_s(T)\,:=\,\<\hat{H}_s\>_{\beta}} &=& {\textstyle \frac{M \hbar}{2 \pi}\,\int_0^{\infty} d\omega\,
    (\omega_0^2 + \omega^2)\,
    \coth\left(\frac{\beta \hbar \omega}{2}\right)}\n\\
    && {\textstyle \times\, \mbox{Im}\{\tilde{\chi}(\omega + i\,0^+)\}\,.}
\end{eqnarray}
In comparison, the internal energy of an uncoupled ({\em or} free)
oscillator is \cite{CAL85}
\begin{equation}\label{eq:internal_energy1}
    e(\omega_0,T)\; =\; \hbar \omega_0 \left(\frac{1}{2}\, +\,
    \<\hat{n}\>_{\beta}\right)\; =\; \frac{\hbar \omega_0}{2} \coth \frac{\beta \hbar
    \omega_0}{2}\,,
\end{equation}
where the average quantum number $\<\hat{n}\>_{\beta} = 1/(e^{\beta
\hbar \omega_0} - 1)$. Its classical counterpart appears as
$e_{cl}(T) = \frac{1}{\beta}$. With (\ref{eq:internal_energy1}),
equation (\ref{eq:energy1}) can now be transformed to an expression
\begin{equation}\label{eq:energy2}
    E_s(T)\; =\; E_s(0)\, +\, \Delta E_s(T)\,,
\end{equation}
where
\begin{eqnarray}
    E_s(0) &=& -\frac{M \hbar}{4\pi i}\,\oint\,d\omega\,\left(\omega_0^2\,+\,\omega^2\right)\, \tilde{\chi}(\omega)\n\\
    \Delta E_s(T) &=& -\frac{M \hbar}{2\pi i}\,\oint\,d\omega\,\left(\omega_0^2\,+\,\omega^2\right)\, \tilde{\chi}(\omega)\,
    \<\hat{n}\>_{\beta}\,.
\end{eqnarray}
Here, the integration path is a loop around the positive real axis
in the complex $\omega$-plane, consisting of the two branches,
$(\infty + i \epsilon,\,i \epsilon)$ and $(-i \epsilon,\,\infty - i
\epsilon)$ \cite{KAM04}. Therefore, $E_s(T)$ for the discrete bath
modes can be exactly obtained in closed form from the residues
evaluated at all poles $\{\bar{\omega}_k\}$ of
$\tilde{\chi}(\omega)$ in (\ref{eq:susceptibility2}) on the positive
real axis. Then, equation (\ref{eq:energy2}) reduces to
\begin{equation}\label{eq:energy_general_treatment1}
    E_s(T)\, =\, \frac{1}{2} \sum_{k=0}^N e(\bar{\omega}_k,T) \left\{1 + \left(\frac{\omega_0}{\bar{\omega}_k}\right)^2\right\}
    \frac{\displaystyle \prod_{j=1}^N\,(\bar{\omega}_k^2 - \omega_j^2)}{\displaystyle \prod_{k'=0 \atop (\ne k)}^N\,(\bar{\omega}_k^2 -
    \bar{\omega}_{k'}^2)}\,.
\end{equation}
To study the quantum second law below, we need two different
Helmholtz free energies for the oscillator coupled to a bath.

We first consider the Helmholtz free energy of the coupled
oscillator, $F_s(T) = E_s(T) - T S_s$ \cite{CAL85} with its entropy
$S_s = -\partial F_s/\partial T$. This can easily be solved for
$F_s$ such that \cite{POL02}
\begin{equation}\label{eq:free_energy_of_system1}
    {\textstyle F_s(T)\; =\; T \left(-\int^T_{T_0} \frac{E_s(T')}{{T'}^2}\,dT'\,
    +\, {\mathcal C}\right)\,.}
\end{equation}
By requiring that the entropy
\begin{equation}\label{eq:entropie_of_system1}
    S_s(\beta)\; =\; k_B\,\beta\,E_s(\beta)\, -\, k_B \int^{\beta}_{\beta_0} E_s(\beta')\,d\beta'\, -\, {\mathcal C}
\end{equation}
with $\beta_0 = 1/k_B T_0$ vanish at zero temperature, we can
determine the constant ${\mathcal C}$ of integration; using
(\ref{eq:energy1}) we easily obtain
\begin{equation}\label{eq:entropy_zero_at_zero1}
    {\mathcal C}/k_B\; =\; {\textstyle \frac{M \hbar}{2 \pi}\,\int_0^{\infty} d\omega\,
    (\omega_0^2 + \omega^2)\, \mbox{Im}\{\tilde{\chi}(\omega + i\,0^+)\}\, {\mathcal A}(\omega)\,,}
\end{equation}
where
\begin{equation}\label{eq:entropy_zero_at_zero2}
    {\textstyle {\mathcal A}(\omega)\; =\;
    \left.\left(\beta \coth \frac{\beta \hbar
    \omega}{2}\, -\, \int_{\beta_0}^{\beta} d\beta' \coth \frac{\beta' \hbar
    \omega}{2}\right)\right|_{\beta \to \infty}\,.}
\end{equation}
The asymptotic series, $\coth z = 1 + 2 \sum_{l=1}^{\infty} e^{-2 l
z}$ then allows equation (\ref{eq:entropy_zero_at_zero2}) to become
\begin{equation}\label{eq:entropy_zero_at_zero3}
    {\textstyle {\mathcal A}(\omega) \; =\; \int^{\beta_0} d\beta' \coth
    \frac{\beta' \hbar \omega}{2}\; =\; \frac{2}{\hbar
    \omega}\,\ln\left(\sinh \frac{\beta_0 \hbar
    \omega}{2}\right)\,,}
\end{equation}
which clearly makes equations (\ref{eq:free_energy_of_system1}) and
(\ref{eq:entropie_of_system1}), respectively, independent of $T_0$
({\em or} $\beta_0$) and accordingly uniquely determined.

The other Helmholtz free energy ${\mathcal F}_s(T)$ needed for the
second law is described as follows; the minimum work required to
couple a system oscillator at temperature $T$ to a bath at the same
temperature is equivalent to the Helmholtz free energy of the {\em
coupled} total system minus the free energy of the {\em uncoupled}
bath \cite{FOR85}. This minimum work can then be obtained as the
free energy ${\mathcal F}_s(T) = -\frac{1}{\beta}\,\ln {\mathcal
Z}_{\beta}$, where the canonical partition function ${\mathcal
Z}_{\beta} = \mbox{Tr}\, e^{-\beta \hat{H}}/\mbox{Tr}_b\, e^{-\beta
\hat{H}_b}$. Here, $\mbox{Tr}_b$ denotes the partial trace for the
bath alone (in the absence of a coupling between system and bath,
this would exactly correspond to the partition function of the
system only). By means of the normal-mode frequencies
$\bar{\omega}_k$, we easily get
\begin{equation}\label{eq:partion_function1}
    {\mathcal Z}_{\beta}\; =\; \frac{\displaystyle
    \prod_{k=0}^N\, \sum_{n_k=0}\, e^{-\beta \hbar \bar{\omega}_k
    \left(n_k + \frac{1}{2}\right)}}{\displaystyle \prod_{j=1}^N\, \sum_{n_j=0}\, e^{-\beta \hbar
    \omega_j \left(n_j + \frac{1}{2}\right)}}\,,
\end{equation}
which yields
\begin{equation}\label{eq:helmholtz_energy0}
    {\mathcal F}_s(T)\; =\; \sum_{k=0}\,f(\bar{\omega}_k,T)\, -\,
    \sum_{j=1}\,f(\omega_j,T)
\end{equation}
with the free energy of an uncoupled oscillator
\begin{equation}\label{eq:free_energy_finite_temp2}
    f(\omega, T)\; =\; \frac{\hbar \omega}{2}\, +\, \frac{1}{\beta}\, \ln\left(1 -
    e^{-\beta \hbar \omega}\right)\,.
\end{equation}
The classical counterpart of (\ref{eq:free_energy_finite_temp2}) is
$f_{cl}(\omega, T) = \frac{\ln \beta \hbar \omega}{\beta}$ with
$\hbar \ll 1$. Equation (\ref{eq:helmholtz_energy0}) can then be
rewritten as \cite{FOR85}
\begin{equation}\label{eq:free_energy_finite_temp1}
    {\mathcal F}_s(T)\; =\; \frac{1}{\pi} \int_0^{\infty} d \omega\, f(\omega,T)\;
    \mbox{Im}\left\{\frac{d}{d\omega}\,\ln \tilde{\chi}(\omega + i
    0^+)\right\}
\end{equation}
in terms of the susceptibility. We note here that for an uncoupled
oscillator, $\mbox{Im}\left\{\frac{d}{d\omega}\,\ln
\tilde{\chi}(\omega + i 0^+)\right\} \to
\pi\,\delta(\omega-\omega_0)$ and thus ${\mathcal F}_s(T) \to
f(\omega_0,T)$. Similarly to (\ref{eq:free_energy_finite_temp1}), we
can also obtain the energy required to couple a system oscillator to
a bath,
\begin{eqnarray}\label{eq:energy_finite_temp1}
    {\mathcal E}_s(T) &=& \sum_{k=0}\,e(\bar{\omega}_k,T)\, -\,
    \sum_{j=1}\,e(\omega_j,T)\\
    &=& \frac{1}{\pi} \int_0^{\infty} d \omega\, e(\omega,T)\;
    \mbox{Im}\left\{\frac{d}{d\omega}\,\ln \tilde{\chi}(\omega + i
    0^+)\right\}\,.\n
\end{eqnarray}
From $e(\omega, T) \geq f(\omega, T)$ (the equal sign holds for $T =
0$ only) with the frequency relationship in
(\ref{eq:frequency_relation3}), we can easily get ${\mathcal E}_s(T)
\geq {\mathcal F}_s(T)$.

We now consider a cyclic process composed of the coupling of a
harmonic oscillator to a bath and then the decoupling of the
oscillator from the bath (the coupling constants $c_j \to 0$). The
free energy change on completion of the coupling process is
${\mathcal F}_s(T) - f(\omega_0,T)$, whereas the maximum useful work
obtainable from the oscillator only in the decoupling process is the
free energy difference $F_s(T) - f(\omega_0,T)$ which cannot be
greater than the energy change $E_s(T) - e(\omega_0,T)$. Here it is
assumed obviously that the extraction of energy from the bath is
impossible. The second law can then be expressed as an inequality
\begin{equation}\label{eq:second_law_finite1}
    {\mathcal F}_s(T)\, -\, f(\omega_0,T)\; \geq\; E_s(T)\, -\, e(\omega_0,T)\,.
\end{equation}
In obtaining (\ref{eq:second_law_finite1}) we used the conceptional
difference between ${\mathcal F}_s(T)$ and $F_s(T)$ (``operational
asymmetry'') \cite{KIM07}. For zero temperature, this inequality,
obviously, reduces to ${\mathcal F}_s(0) \geq E_s(0)$, the validity
of which has been explicitly proven for the Drude damping model
\cite{FOR06} and for the discrete bath modes, by means of $E_s(0)$
in (\ref{eq:energy_general_treatment1}) and $\mathcal{F}_s(0)$ in
(\ref{eq:helmholtz_energy0}) with the frequency relationship
(\ref{eq:frequency_relation3}), and the generalized realistic
damping models of continuous bath modes \cite{KIM06}. For non-zero
temperatures, on the other hand, it is very non-trivial to
investigate the validity of inequality (\ref{eq:second_law_finite1})
with $E_s(T)$ and $\mathcal{F}_s(T)$ for the discrete bath modes.
For the continuous bath modes, the evaluation of $E_s(T)$ and
${\mathcal F}_s(T)$ clearly depends on the parameters of the damping
model considered. We will below discuss inequality
(\ref{eq:second_law_finite1}) explicitly within the Drude model
which is the prototype for physically realistic damping.

\section{Discussion of the second law within the Drude model}\label{sec:2nd_law_drude}
It is convenient in the Drude model to adopt, in place of
$(\omega_0, \omega_d, \gamma_o)$, the parameters $({\mathbf w}_0,
\Omega, \gamma)$ through the relations \cite{FOR06}
\begin{eqnarray}\label{eq:parameter_change0}
    &\omega_0^2\; :=\; {\mathbf w}_0^2\; \frac{\Omega}{\Omega\, +\, \gamma}\,;\;
    \omega_d\; :=\; \Omega\, +\, \gamma&\n\\
    &\gamma_o\; :=\; \gamma\, \frac{\Omega\, (\Omega\, +\, \gamma)\,
    +\, {\mathbf w}_0^2}{(\Omega\, +\, \gamma)^2}\,.&
\end{eqnarray}
Substituting equation (\ref{eq:drude_gamma1}) with
(\ref{eq:parameter_change0}) into (\ref{eq:chi_tilde1}), we obtain
the susceptibility
\begin{equation}\label{eq:susceptibility_drude1}
    \tilde{\chi}_d(\omega)\; =\; -\frac{1}{M}\,
    \frac{\omega\, +\, i\,(\Omega\, +\, z_1\, +\, z_2)}{(\omega\, +\, i \Omega)
    (\omega\, +\, i z_1) (\omega\, +\, i z_2)}\,,
\end{equation}
where $z_1 = \gamma/2 + i {\mathbf w}_1$ and $z_2 = \gamma/2 - i
{\mathbf w}_1$ with ${\mathbf w}_1 = \sqrt{{\mathbf w}_0^2 -
(\gamma/2)^2}$. First, we consider the overdamped case $(\gamma/2
> {\mathbf w}_0)$, where $z_1,\,z_2 > 0$. We then have
\begin{equation}\label{eq:im_chi1}
    \mbox{Im}\,\tilde{\chi}_d(\omega)\; =\; -\frac{1}{M}\,\sum_{l=1}^3\,\lambda_d^{(l)}\, \frac{\omega}{\omega^2 +
    \underline{\omega_l}^2}\,,
\end{equation}
where $\underline{\omega_1} = \Omega$, $\underline{\omega_2} = z_1$,
$\underline{\omega_3} = z_2$, and the coefficients
\begin{eqnarray}\label{eq:coefficients}
    &\lambda_d^{(1)}\; =\; \frac{z_1\,+\,z_2}{(\Omega\,-\,z_1) (z_2\,-\,\Omega)}\;;\;
    \lambda_d^{(2)}\; =\; \frac{\Omega\,+\,z_2}{(z_1\,-\,\Omega) (z_2\,-\,z_1)}&\n\\
    &\lambda_d^{(3)}\; =\; \frac{\Omega\,+\,z_1}{(z_2\,-\,\Omega) (z_1\,-\,z_2)}\,.&
\end{eqnarray}
Here, we note that
\begin{equation}\label{eq:drude_coefficient_relations}
    \sum_{l=1}^3\,\lambda_d^{(l)}\; =\; 0\;\; ;\;\; \sum_{l=1}^3\,\lambda_d^{(l)}\,\underline{\omega_l}^2\; =\; 0\,.
\end{equation}

To obtain an explicit expression for the energy $E_s^{(d)}(T)$ of
the coupled oscillator, we first consider the integral in
(\ref{eq:dissipation_fluctuation_theorem}); by performing a contour
integration with the aid of (\ref{eq:im_chi1}) and the identity
\begin{equation}\label{eq:hyperbolic_cot}
    \coth\left(\frac{\beta \hbar \omega}{2}\right)\; =\;
    \frac{2}{\beta \hbar \omega} \left(1\, +\, 2
    \sum_{n=1}^{\infty}\, \frac{\omega^2}{\nu_n^2 +
    \omega^2}\right)\,,
\end{equation}
where $\nu_n = 2\pi n/\beta \hbar$, we can have
\begin{eqnarray}\label{eq:x_squared1}
     \hspace*{-.5cm}&& \textstyle \frac{1}{2}\, \left\<\hat{q}(0)\,\hat{q}(t)\, +\,
     \hat{q}(t)\,\hat{q}(0)\right\>_{\beta}^{(d)}\; =\\
     \hspace*{-.5cm}&& {\textstyle -\frac{1}{\beta M}}
     \sum_{l=1}^3\,\lambda_d^{(l)}\,\left\{{\textstyle \frac{e^{-\underline{\omega_l} t}}{\underline{\omega_l}}}\,
     +\, 2\,\sum_{n=1}^{\infty}\,{\textstyle \frac{\nu_n\,e^{-\nu_n t}\, -\,
     \underline{\omega_l}\,e^{-\underline{\omega_l} t}}{\nu_n^2\, -\, \underline{\omega_l}^2}}\right\}\,.\n
\end{eqnarray}
With $\hbar \to 0$, this reduces to its classical counterpart,
$_{cl}\hspace*{-0.03cm}\<q(0)\,q(t)\>_{\beta}^{(d)} =
-\frac{1}{\beta M}
\sum_l\,\lambda_d^{(l)}\,\frac{e^{-\underline{\omega_l}
t}}{\underline{\omega_l}}$. From (\ref{eq:x_squared1}) and the
relation \cite{ING98}
\begin{equation}\label{eq:x_squared2}
    \textstyle \<\dot{\hat{q}}(0)\,\dot{\hat{q}}(t)\, +\,
    \dot{\hat{q}}(t)\,\dot{\hat{q}}(0)\>_{\beta}\, =\, -\frac{d^2}{d t^2}\,\<\hat{q}(0)\,\hat{q}(t)\, +\,
    \hat{q}(t)\,\hat{q}(0)\>_{\beta}
\end{equation}
it can eventually be found that
\begin{eqnarray}\label{eq:energy_drude_overdamped1}
    E_s^{(d)}(T) &=& {\textstyle \frac{1}{\beta}}
    \sum_{l=1}^3\,\lambda_d^{(l)}\, \times\\
    && \left\{{\textstyle \frac{\omega_0^2\,-\,\underline{\omega_l}^2}{2\,\underline{\omega_l}}}\,
    -\, \sum_{n=0}^{\infty}\,{\textstyle \frac{\omega_0^2\, -\,
    \left(\nu_n^2\,+\,\nu_n\,\underline{\omega_l}\,+\,\underline{\omega_l}^2\right)}{\nu_n\, +\, \underline{\omega_l}}}\right\}\,.\n
\end{eqnarray}
With $\hbar \to 0$, its classical counterpart easily appears as
$_{cl}\hspace*{-0.03cm}E_s^{(d)}(T) = \frac{1}{2 \beta} \sum_l
\lambda_d^{(l)}\,\frac{\underline{\omega_l}^2 -
\omega_0^2}{\underline{\omega_l}} = \frac{1}{\beta} =
e_{cl}(\omega_0,T)$. We can also express (\ref{eq:x_squared1}) and
(\ref{eq:x_squared2}) at $t=0$, respectively, in terms of the
Digamma function \cite{ABS74}
\begin{equation}
    \psi(y)\; =\; \frac{d\,\ln \Gamma(y)}{d y}\; =\; -c_e\, +\, \sum_{n=1}^{\infty}\,\frac{1}{n}\, -\, \sum_{n=0}^{\infty}\,\frac{1}{n +
    y}
\end{equation}
with the Euler constant $c_e$ as
\begin{eqnarray}
    \hspace*{-.5cm}\<\hat{q}^2\>_{\beta}^{(d)} &=& {\textstyle \frac{1}{M}}
    \sum_{l=1}^3 {\textstyle \lambda_d^{(l)}\,\left\{\frac{1}{\beta \underline{\omega_l}}\,
    +\, \frac{\hbar}{\pi}\; \psi\left(\frac{\beta \hbar \underline{\omega_l}}{2 \pi}\right)\right\}}\label{eq:x_x_dot_drude_overdamped1}\\
    \hspace*{-.5cm}\<\dot{\hat{q}}^2\>_{\beta}^{(d)} &=& {\textstyle -\frac{1}{M}}
    \sum_{l=1}^3 {\textstyle \lambda_d^{(l)}\,\underline{\omega_l}^2\,\left\{\frac{1}{\beta \underline{\omega_l}}\,
    +\, \frac{\hbar}{\pi}\; \psi\left(\frac{\beta \hbar \underline{\omega_l}}{2 \pi}\right)\right\}}\,.\label{eq:x_x_dot_drude_overdamped2}
\end{eqnarray}
Here, we used (\ref{eq:drude_coefficient_relations}). Equation
(\ref{eq:energy_drude_overdamped1}) can thus be rewritten as
\begin{equation}\label{eq:energy_drude_overdamped2}
    E_s^{(d)}(T)\; =\; {\textstyle \frac{1}{2}}
    \sum_{l=1}^3 {\textstyle \lambda_d^{(l)}\,(\omega_0^2\,-\,\underline{\omega_l}^2)\, \left\{\frac{1}{\beta \underline{\omega_l}}\,
    +\, \frac{\hbar}{\pi}\; \psi\left(\frac{\beta \hbar \underline{\omega_l}}{2
    \pi}\right)\right\}}\,.
\end{equation}
With the aid of the asymptotic expression, $\psi(y) = \ln y -
\frac{1}{2 y} - \sum^{\infty}_{n=1} \frac{B_{2n}}{2n\,y^{2n}}$ with
the Bernoulli number $B_n$ \cite{ABS74}, the zero-temperature value
$E_s^{(d)}(0)$, clearly, reduces to (\ref{eq:energy_drude1}) derived
in \cite{KIM06} (see Appendix).

Let us now consider the free energy ${\mathcal F}_s^{(d)}(T)$. By
substituting (\ref{eq:susceptibility_drude1}) into
(\ref{eq:free_energy_finite_temp1}) with the identity, $\ln (1 + y)
= -\sum_{n=1}^{\infty}\,(-y)^{n}/n$, we can easily obtain
\begin{equation}\label{eq:free_energy_finite_temp4}
    {\mathcal F}_s^{(d)}(T)\; =\; {\mathcal F}_s^{(d)}(0)\, +\, \frac{1}{\pi\beta}\,
    \sum_{n=1}^{\infty}\,\frac{1}{n}\, \Delta_n(\beta)\,,
\end{equation}
where
\begin{eqnarray}\label{eq:helmholtz_energy}
    {\mathcal F}_s^{(d)}(0) &=&
    \frac{\hbar}{2 \pi}\, \left\{(\Omega + \gamma)\,
    \ln\left(\frac{\Omega + \gamma}{\Omega}\right)\,
    +\, \gamma\, \ln\left(\frac{\Omega}{{\mathbf w}_0}\right)\, +\right.\n\\
    && \left.\bar{{\mathbf w}}_1\, \ln \left(\frac{\gamma/2\, -\, \bar{{\mathbf w}}_1}{\gamma/2\,
    +\, \bar{{\mathbf w}}_1}\right)\right\}
\end{eqnarray}
with $\bar{{\mathbf w}}_1 = \sqrt{(\gamma/2)^2 - {\mathbf w}_0^2}$,
as was derived in \cite{KIM06}, and
\begin{eqnarray}
    \Delta_n(\beta) &=& \sum_{\mu=0}^{3}\, \tau_d^{(\mu)} \int_0^{\infty} d y\,
    \frac{e^{-n \beta \hbar \underline{\omega_{\mu}}\,y}}{y^2 + 1}\label{eq:free_energy_finite_temp5}\\
    &=& \sum_{\mu=0}^{3}\, \textstyle \tau_d^{(\mu)}\, \left\{\sin(n \beta \hbar \underline{\omega_{\mu}})\;
    \mbox{Ci}(n \beta \hbar \underline{\omega_{\mu}})\; -\right.\n\\
    && \left.\cos(n \beta \hbar \underline{\omega_{\mu}})\; \mbox{si}(n \beta \hbar
    \underline{\omega_{\mu}})\right\}\label{eq:free_energy_finite_temp6}
\end{eqnarray}
with $\tau_d^{(0)} = 1$, $\tau_d^{(1)} = \tau_d^{(2)} = \tau_d^{(3)}
= -1$, and $\underline{\omega_0} = \omega_d$ (see also Appendix). By
the substituting the classical quantity $f_{cl}(\omega,T)$ into
(\ref{eq:free_energy_finite_temp1}) with $\int_0^x dy/(y^2 + 1) =
\arctan x$ and $\int_0^{\infty} dy \ln y/(y^2 + 1) = 0$
\cite{GRA00}, we can also obtain the classical counterpart
$_{cl}\hspace*{-0.03cm}{\mathcal F}_s^{(d)}(T) = -\frac{1}{2 \beta}
\ln \frac{\omega_d}{\Omega} + f_{cl}({\mathbf w}_0,T) =
f_{cl}(\omega_0,T)$.

For the underdamped case $(\gamma/2 \leq {\mathbf w}_0)$, equations
(\ref{eq:energy_drude_overdamped2}) and
(\ref{eq:free_energy_finite_temp4}) can be found to hold as well,
respectively, being expressed in terms of the functions with
complex-valued arguments. By showing the validity of inequality
(\ref{eq:second_law_finite1}) for underdamped and overdamped cases,
\begin{equation}\label{eq:K_d1}
    K_d(T)\, :=\, {\mathcal F}_s^{(d)}(T) - f(\omega_0,T) - E_s^{(d)}(T) + e(\omega_0,T)\, \geq\, 0
\end{equation}
as in Fig. \ref{fig:figure}, we see that there is no violation of
the quantum second law; in fact, $K_d(T)$ vanishes asymptotically
with the increase of $T$.

Comments deserve here. In the classical treatment both sides of
inequality (\ref{eq:second_law_finite1}) vanish, namely,
$_{cl}\hspace*{-0.03cm}K_d(T) = 0$, which clearly means that no work
is required to couple a linear system to a bath at the same
temperature, and no energy change in the system is obtained during
the decoupling (and the coupling). In the quantum treatment, on the
other hand, $E_s^{(d)}(T)$ and ${\mathcal F}_s^{(d)}(T)$ depend on
the damping parameters, respectively. Therefore, while both sides of
(\ref{eq:second_law_finite1}) become vanishing in the high
temperature limit (equivalently, $\hbar \to 0$), they actually do
not vanish especially in the low temperature regime. This
non-vanishing behavior stems from the system-bath entanglement
induced by the finite coupling strength between them, which leads to
the deviation from $\hat{\rho}_{\beta}^{(s)} = e^{-\beta
\hat{H}_s}/Z_{\beta}^{(s)}$ for the reduced density matrix
$\hat{\rho}_s^{(d)}(T)$ being, clearly, damping-parameter dependent.

In fact, we have $K_d(T) > 0$ especially in the low temperature
regime (see Fig.~\ref{fig:figure}). This strict irreversibility over
a single cycle composed of the coupling and decoupling process
appears from the fact that the system-bath entanglement induces the
entanglement between any pair of infinitely many bath oscillators
(``entanglement swapping'' \cite{ALB01}), which cannot completely
removed over the system-bath decoupling process. Therefore, we
essentially cannot recover the original state of the bath,
$\hat{\rho}_{\beta}^{(b)}$ and thus that of the system,
$\hat{\rho}_{\beta}^{(s)}$. As a result, ${\mathcal F}_s^{(d)}(T) -
f(\omega_0,T)$, being the minimum work required for the entangling
in the coupling process, is greater than the energy change
$E_s^{(d)}(T) - e(\omega_0,T)$, which can necessarily not be less
than the free energy change, $F_s^{(d)}(T) - f(\omega_0,T)$ being
the maximum useful work obtainable from the system only in the
decoupling process. With the increase of $T$, however, the strict
irreversibility shrinks ($K_d(T) \to 0^+$) since the thermal effect
dominates the quantum effect. In the classical case, on the other
hand, this operational asymmetry, introduced in the last paragraph
of Sect. \ref{sec:2nd_law}, disappears at an arbitrary temperature,
namely $_{cl}\hspace*{-0.03cm}{\mathcal F}_s^{(d)}(T)
=\,_{cl}\hspace*{-0.03cm}F_s^{(d)}(T) = f_{cl}(\omega_0,T)$.

\section{Comparison with the Ohmic model}\label{sec:ohmic}
Let us briefly consider the Ohmic model (as a cutoff-free damping
model) for an arbitrary temperature to compare with the Drude model
considered in Sect. \ref{sec:2nd_law_drude}. For zero temperature it
is known \cite{KIM06} that $K_o(0) = {\mathcal F}_s^{(o)}(0) -
E_s^{(o)}(0)$ vanishes, where
\begin{equation}
    \textstyle {\mathcal F}_s^{(o)}(0)\, =\, E_s^{(o)}(0)\, =\,
    \frac{\hbar \gamma_o}{2 \pi} \int_0^{\infty}
    d\omega\,\frac{\omega\,\left(\omega^2\,+\,\omega_0^2\right)}{\left(\omega^2\,-\,\omega_0^2\right)^2\,+\,\left(\gamma_0 \omega\right)^2}
\end{equation}
diverges logarithmically, while its Drude-model counterpart $K_d(0)
\to E_g\,\gamma/\pi {\mathbf w}_0$ in the limit $\omega_d \to
\infty$ (equivalently, $\Omega \to \infty$) where $E_g$ is the
ground state energy of a free oscillator.

For the overdamped case $(\gamma_o/2
> \omega_0)$, the susceptibility in (\ref{eq:chi_tilde1}) appears as
\begin{equation}\label{eq:ohmic1}
    \tilde{\chi}_o(\omega)\; =\; -\frac{1}{M}\,\frac{1}{(\omega + i \omega_1)\,(\omega + i \omega_2)}\,,
\end{equation}
where $\omega_1 = \gamma_o/2 - \bar{\mathbf w} > 0$ and $\omega_2 =
\gamma_o/2 + \bar{\mathbf w} > 0$ with $\bar{\mathbf w} =
\sqrt{(\gamma_o/2)^2 - \omega_0^2}$. Substituting (\ref{eq:ohmic1})
into (\ref{eq:dissipation_fluctuation_theorem}), we can obtain
\begin{eqnarray}\label{eq:x_squared_ohm1}
     \hspace*{-.5cm}&& \textstyle \frac{1}{2}\, \left\<\hat{q}(0)\,\hat{q}(t)\, +\,
     \hat{q}(t)\,\hat{q}(0)\right\>_{\beta}^{(o)}\; =\\
     \hspace*{-.5cm}&& {\textstyle -\frac{1}{2 \beta \bar{\mathbf w} M}}
     \sum_{j=1}^2\,\lambda_o^{(j)}\,\left\{{\textstyle \frac{e^{-\omega_j t}}{\omega_j}}\,
     +\, 2\,\sum_{n=1}^{\infty}\,{\textstyle \frac{\nu_n\,e^{-\nu_n t}\, -\,
     \omega_j\,e^{-\omega_j t}}{\nu_n^2\, -\, \omega_j^2}}\right\}\n
\end{eqnarray}
with $\lambda_o^{(1)} = -1$ and $\lambda_o^{(2)} = 1$ \cite{ILK07},
from which, similarly to (\ref{eq:x_x_dot_drude_overdamped1}),
\begin{equation}\label{eq:x_x_ohm_overdamped1}
    \<\hat{q}^2\>_{\beta}^{(o)}\; =\; {\textstyle \frac{1}{2 \bar{\mathbf w} M}}
    \sum_{j=1}^2 {\textstyle \lambda_o^{(j)}\,\left\{\frac{1}{\beta\,\omega_j}\,
    +\, \frac{\hbar}{\pi}\; \psi\left(\frac{\beta \hbar \omega_j}{2 \pi}\right)\right\}}\,.
\end{equation}
Substituting (\ref{eq:x_squared_ohm1}) into (\ref{eq:x_squared2}),
we can also get
\begin{eqnarray}\label{eq:x_x_dot_ohm_overdamped1}
    \<\dot{\hat{q}}^2\>_{\beta}^{(o)} &=& {\textstyle -\frac{1}{2 \bar{\mathbf w} M}}
    \sum_{j=1}^2 {\textstyle \lambda_o^{(j)}\,\omega_j^2\,\left(\frac{1}{\beta \omega_j}\,
    +\, \frac{\hbar}{\pi}\left\{\psi\left(\frac{\beta \hbar \omega_j}{2 \pi}\right)\,
    +\right.\right.}\n\\
    && \left.\left.c_e\, -\, \sum_{n=1}^{\infty} \textstyle
    \frac{1}{n}\right\}\right)\,,
\end{eqnarray}
which diverges logarithmically for an arbitrary temperature (note
that $\lim_{n \to \infty} (\sum_{k=1}^n \frac{1}{k} - \ln n) =
c_e$); compare this with $\<\dot{\hat{q}}^2\>_{\beta}^{(d)}$ in
(\ref{eq:x_x_dot_drude_overdamped2}) being convergent. From
(\ref{eq:x_x_ohm_overdamped1}) and
(\ref{eq:x_x_dot_ohm_overdamped1}), the energy of the coupled
oscillator is
\begin{eqnarray}\label{eq:energy_ohm_overdamped2}
    E_s^{(o)}(T) &=& {\textstyle \frac{1}{4\bar{\mathbf w}}}
    \sum_{j=1}^2 {\textstyle \lambda_o^{(j)} \left\{(\omega_0^2 - \omega_j^2) \left\{\frac{1}{\beta \omega_j} +
    \frac{\hbar}{\pi} \psi\left(\frac{\beta \hbar \omega_j}{2
    \pi}\right)\right\}\right.}\n\\
    && \left.{\textstyle - \frac{\hbar \omega_j^2}{\pi}} \left(c_e\, -\,
    \sum_{n=1}^{\infty} \textstyle \frac{1}{n}\right)\right\}\,,
\end{eqnarray}
which clearly diverges. With $\hbar \to 0$, its classical
counterpart, however, reduces to $_{cl}\hspace*{-0.03cm}E_s^{(o)}(T)
= e_{cl}(\omega_0,T)$ (it can also be found \cite{HAE06} that
$E_s^{(o)}(T)$ is identical to the energy ${\mathcal E}_s^{(o)}(T)$
obtainable from (\ref{eq:energy_finite_temp1})). Further, similarly
to (\ref{eq:free_energy_finite_temp4}), we can easily obtain
\begin{equation}\label{eq:energy_ohm_free_energy}
    {\mathcal F}_s^{(o)}(T)\; =\; {\mathcal F}_s^{(o)}(0)\, +\, \Delta {\mathcal
    F}_s^{(o)}(T)\,,
\end{equation}
where
\begin{eqnarray}\label{eq:energy_ohm_free_energy1}
    \Delta {\mathcal F}_s^{(o)}(T) &=& -\, \frac{1}{\pi \beta}
    \sum_{n=1}^{\infty}\,\frac{1}{n}\,\sum_{j=1}^2\,\{\,\sin(n \beta \hbar \omega_j)\,\mbox{Ci}(n\hbar\beta \omega_j)\n\\
    && -\, \cos(n \beta \hbar \omega_j)\,\mbox{si}(n \beta \hbar \omega_j)\}\,.
\end{eqnarray}
The free energy ${\mathcal F}_s^{(o)}(T)$ also diverges. With $\hbar
\to 0$, we get $_{cl}\hspace*{-0.03cm}{\mathcal F}_s^{(o)}(T) =
f_{cl}(\omega_0,T)$. Equations (\ref{eq:energy_ohm_overdamped2}) and
(\ref{eq:energy_ohm_free_energy}) can be found to hold,
respectively, for the underdamped case as well.

For comparison with the Drude model, we take the limit $\Omega$
({\em or} $\omega_d$) $\to \infty$ in (\ref{eq:coefficients}) so
that $\lambda_d^{(1)} \to 0$, $\lambda_d^{(2)} \to 1/(z_1 - z_2)$,
and $\lambda_d^{(3)} \to -\lambda_d^{(2)}$. Then, it can easily be
shown that $E_s^{(d)}(T) \nrightarrow E_s^{(o)}(T)$ and ${\mathcal
F}_s^{(d)}(T) \nrightarrow {\mathcal F}_s^{(o)}(T)$; the second term
on the right hand side of (\ref{eq:free_energy_finite_temp4})
reduces to $\Delta {\mathcal F}_s^{(o)}(T)$ in
(\ref{eq:energy_ohm_free_energy}), however, the first term
$\mathcal{F}_s^{(d)}(0) \nrightarrow \mathcal{F}_s^{(o)}(0)$ as
discussed in \cite{KIM06}. As a result, the second-law inequality
(\ref{eq:second_law_finite1}) for the Drude model with $\omega_d \to
\infty$ is not equivalent to that for the Ohmic model. Whereas the
classical counterpart $_{cl}\hspace*{-0.03cm}K_o(T) = 0$, both sides
of (\ref{eq:second_law_finite1}) come to diverge differently so that
it is non-trivial to explicitly evaluate $K_o(T)$ for this {\em
unrealistic} damping model.

\section{Conclusions}
In summary, we have studied the second law in the scheme of quantum
Brownian motion at an {\em arbitrary} temperature. It is clearly a
generalization of the previous works for zero temperature by Ford
and O'Connell \cite{FOR06} and by the authors of the present paper
\cite{KIM06}. It has been shown for the physically realistic damping
model that the work needed to couple a system oscillator to a bath
at the same temperature cannot be less than the work obtainable from
the oscillator only when it is extracted from the bath; especially
in the low temperature regime the apparent irreversibility, $K_d(T)
> 0$, stemming from the system-bath entanglement was found, which is
different from the behavior of its classical counterpart,
$_{cl}\hspace*{-0.03cm}K_d(T) = 0$. Therefore, the quantum second
law holds for an arbitrary temperature. The question about the
validity of the quantum second law for a broader class of quantum
systems than the quantum Brownian motion considered here, especially
non-linear systems coupled to a bath, clearly remains open.

\section*{Acknowledgments}
One of us (I. K.) is grateful to Professor E. Merzbacher (UNC-Chapel
Hill), who kindly encouraged him to pay attention to this subject.

\appendix*\section{: Mathematical supplements}\label{sec:appendix}
It has been shown in \cite{KIM06} that for the overdamped case
$({\mathbf w}_0 \leq \gamma/2)$,
\begin{equation}\label{eq:energy_drude1}
    E_s^{(d)}(0)\; =\; \frac{\hbar}{2 \pi}\, \left\{A({\mathbf w}_0, \Omega, \gamma)\,
    +\, B({\mathbf w}_0, \Omega, \gamma)\right\}\,,
\end{equation}
where
\begin{eqnarray}
    && A({\mathbf w}_0, \Omega, \gamma)\;
    =\label{eq:energy_drude2}\\
    && \textstyle \frac{({\mathbf w}_0^2 + \Omega^2)\,
    (\Omega\,\gamma^2/4\, -\, \Omega\,{\mathbf w}_0^2\, -\, {\mathbf w}_0^2\,\gamma/2)\, +\,
    \Omega^2\,\gamma^3/4}{\bar{{\mathbf w}}_1\,
    (\Omega\, +\, \gamma)\, ({\mathbf w}_0^2\, -\, \Omega\,\gamma\,
    +\, \Omega^2)}\, \ln\left(\frac{\gamma/2\, -\, \bar{{\mathbf w}}_1}{\gamma/2\,
    +\, \bar{{\mathbf w}}_1}\right)\n
\end{eqnarray}
with $\bar{{\mathbf w}}_1 = \sqrt{(\gamma/2)^2 - {\mathbf
w}_0^2}$\,, and
\begin{equation}\label{eq:energy_drude3}
    \textstyle B({\mathbf w}_0, \Omega, \gamma)\; =\; \frac{\Omega\,\gamma\, (\Omega^2\, +\,
    \Omega\,\gamma\, -\, {\mathbf w}_0^2)}{(\Omega\, +\, \gamma)\, ({\mathbf w}_0^2\, -\,
    \Omega\,\gamma\, +\, \Omega^2)}\, \ln (\Omega/{\mathbf w}_0)\,.
\end{equation}

In derivation of equation (\ref{eq:free_energy_finite_temp6}) from
(\ref{eq:free_energy_finite_temp5}), we used \cite{GRA00}
\begin{equation}\label{eq:integral_n_1_1}
    \textstyle \int_0^{\infty} dy\, \frac{e^{-a y}}{y^2 + b^2}\, =\,
    \frac{1}{b}\,\{\sin(a b)\; \mbox{Ci}(a b)\, -\, \cos(a b)\; \mbox{si}(a b)\}\,,
\end{equation}
where $a,\,b > 0$; the sine integral\, $\mbox{si}(y) =
-\int_y^{\infty} dz\,\frac{\sin(z)}{z} = -\frac{\pi}{2} +
\mbox{Si}(y)$ with $\mbox{Si}(y) = \int_0^y dz\,\frac{\sin(z)}{z}$,
and the cosine integral $\mbox{Ci}(y) = -\int_y^{\infty}
dz\,\frac{\cos(z)}{z} = c_e + \ln y + \int_0^y dz\,\frac{\cos(z) -
1}{z}$ with the Euler constant $c_e = 0.5772156649 \cdots$.

\begin{figure}
    \hspace*{-3cm}
    \resizebox{0.75\textwidth}{!}{%
    \includegraphics{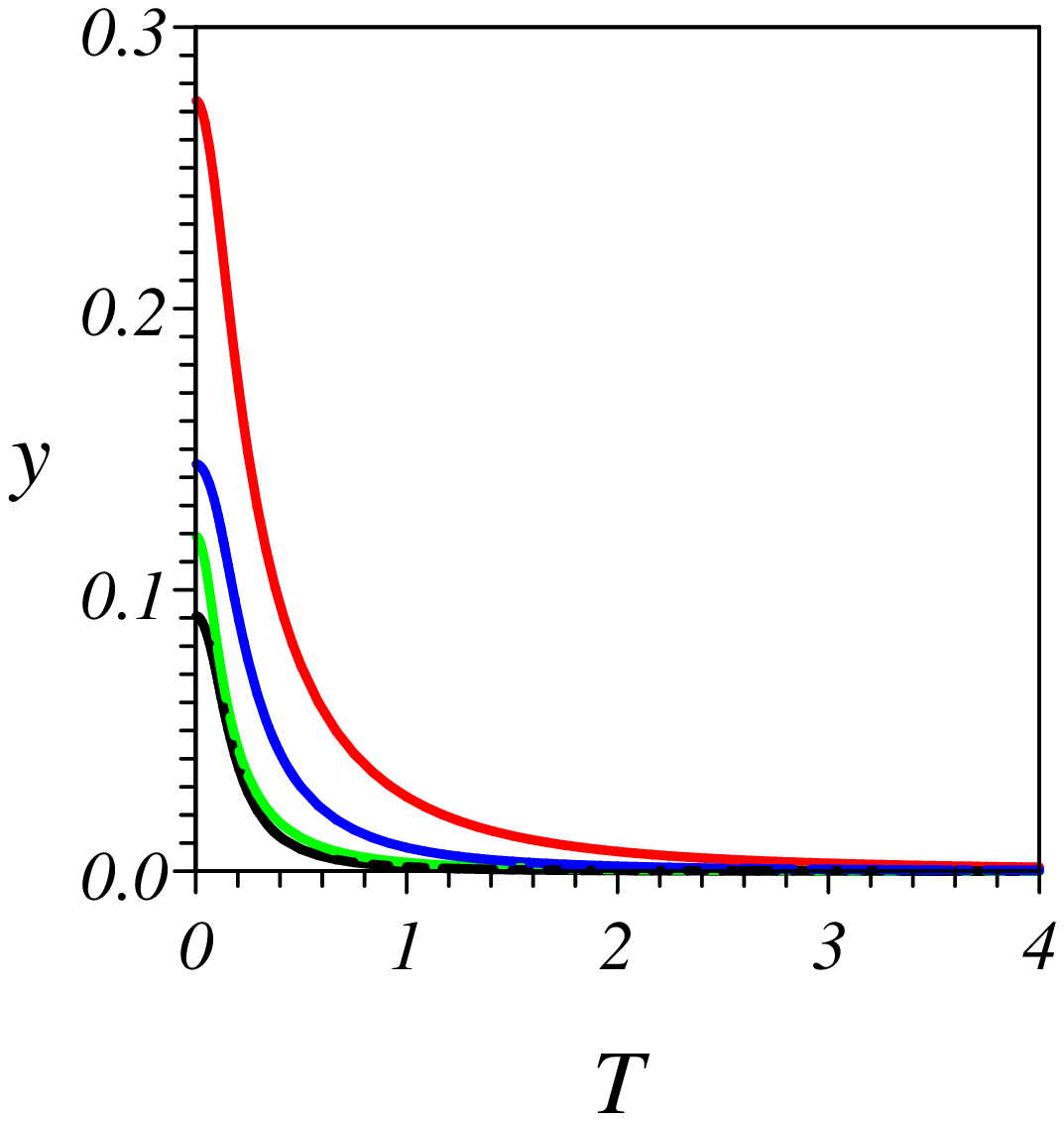}}
    \caption{\label{fig:figure}}
\end{figure}

%
Fig.~\ref{fig:figure}: $y = K_d(T)/\hbar {\mathbf w}_0$ versus
temperature $T$; from bottom to top: ($\Omega=1$ and $\gamma=3/2$
underdamped), ($\Omega=1$ and $\gamma=4$ overdamped), ($\Omega=5$
and $\gamma=3/2$ underdamped), and ($\Omega=5$ and $\gamma=4$
overdamped); here, $\hbar=k_B={\mathbf w}_0=1$; in the high
temperature limit, we have the classical behavior, $y \to 0^+$.
%
%

\begin{thebibliography}{1}
%
\bibitem{FOR06} G.W. Ford and R.F. O'Connell, Phys. Rev. Lett. {\bf 96}, 020402 (2006).
%
\bibitem{KIM06} I. Kim and G. Mahler, Eur. Phys. J. B {\bf 54}, 405 (2006).
%
\bibitem{CAL85} H.B. Callen, {\em Thermodynamics and an
introduction to thermostatics}, 2nd edn. (John Wiley, 1985).
%
\bibitem{CAP05} C. Vladislav and D.P. Sheehan,
{\em Challenges to the second law of thermodynamics: theory and
experiment} (Springer, New York, 2005); {\em Quantum limits to the
second law}, edited by D.P. Sheehan, AIP Conference Proceedings, No.
{\bf 643} (2002).
%
\bibitem{SPI05} V. \v{S}pi\v{c}ka, Th.M. Nieuwenhuizen, and P.D.
Keefe, Physica E {\bf 29}, 1 (2005) and references therein.
%
\bibitem{MAH04} J. Gemmer, M. Michel, and G. Mahler, {\em Quantum Thermodynamics} (Springer, Berlin, 2004).
%
\bibitem{FOR05} G.W. Ford and R.F. O'Connell, Physica E {\bf 29}, 82 (2005).
%
\bibitem{BUT05} M. B\"{u}ttiker and A.N. Jordan, Physica E {\bf 29}, 272 (2005).
%
\bibitem{HAE06} P. H\"anggi and G.-L. Ingold, Acta Physica Polonica B {\bf 37}, 1537 (2006).
%
\bibitem{SHE05} D.P. Sheehan, J.H. Wright, A.R. Putnam, and E.K.
Perttu, Physica E {\bf 29}, 87 (2005) and references therein.
%
\bibitem{BER05} J. Berger, Physica E {\bf 29}, 100 (2005).
%
\bibitem{FOR85} G.W. Ford, J.T. Lewis, and R.F. O'Connell, Phys. Rev. Lett. {\bf 55}, 2273 (1985).
%
\bibitem{FOR88} G.W. Ford, J.T. Lewis, and R.F. O'Connell, Phys. Rev. A {\bf 37}, 4419 (1988).
%
\bibitem{LEW88} G.W. Ford, J.T. Lewis, and R.F. O'Connell, J. Stat. Phys. {\bf 53}, 439 (1988).
%
\bibitem{HAE05} P. H\"anggi and G.-L. Ingold, Chaos {\bf 15}, 026105 (2005).
%
\bibitem{GHO05} P.K. Ghosh, D. Barik, and D.S. Ray, Phys. Rev. E {\bf 71}, 041107 (2005).
%
\bibitem{ALL00} A.E. Allahverdyan and Th.M. Nieuwenhuizen, Phys.
Rev. Lett. {\bf 85}, 1799 (2000).
%
\bibitem{NIE02} Th.M. Nieuwenhuizen and A.E. Allahverdyan, Phys.
Rev. E {\bf 66}, 036102 (2002).
%
\bibitem{WEI99} U. Weiss, {\em Quantum dissipative systems}, 2nd edn.
(World Scientific, Singapore, 1999).
%
\bibitem{ING98} G.-L. Ingold, {\em Dissipative quantum systems} in
{\em Quantum transport and dissipation} (Wiley-VCH, 1998), pp
213-248.
%
\bibitem{ZWA01} R. Zwanzig, {\em Nonequilibrium statistical mechanics} (Oxford University Press, 2001)
%
\bibitem{SEN95} I.R. Senitzky, Phys. Rev. E {\bf 51}, 5166 (1995).
%
\bibitem{KIM04} I. Kim and G.J. Iafrate, Found. Phys. Lett. {\bf 17}, 507 (2004).
%
\bibitem{KAM04} N.G. van Kampen, J. Stat. Phys. {\bf 115}, 1057 (2004).
%
\bibitem{POL02} A.D. Polyanin and V.F. Zaitsev, {\em Handbook of Exact Solutions for Ordinary Differential
Equations}, 2nd edn. (CRC Press, New York, 2002).
%
\bibitem{KIM07} In the concluding part of \cite{FOR06} it was argued
that ``${\mathcal F}_s(T)$ is the minimum work required to couple
the atom to the bath while it is also the maximum work that can be
obtained when the atom is extracted from the bath.'' This would,
however, be misleading since ${\mathcal F}_s(T)$ was incorrectly
identified to $F_s(T)$.
%
\bibitem{ABS74} M. Abramowitz and I. Stegun, {\em Handbook of Mathematical Functions
with Formulas, Graphs, and Mathematical Tables} (Dover, New York,
1974).
%
\bibitem{GRA00} I.S. Gradshteyn and I.M. Ryzhik,
{\em Table of Integrals, Series, and Products}, 6th edn. (Academic
Press, San Diego, 2000).
%
\bibitem{ALB01} G. Alber, T. Beth, M. Horodecki, et al.,
{\em Quantum Information}: {\em An introduction to basic theoretical
concepts and experiments} (Sprnger, Berlin, 2001).
%
\bibitem{ILK07} With the aid of (\ref{eq:hyperbolic_cot}) and $\coth z = \frac{\sinh 2x - i \sin
2y}{\cosh 2x - \cos 2y}$ for $z = x + i y$, equation
(\ref{eq:x_squared_ohm1}) can exactly be transformed into equation
(4.87) in \cite{ING98} for the underdamped case.
%
\end{thebibliography}
\end{document}